\documentstyle[preprint,epsf]{jpsj}
\def\runtitle{Spin Defects in Spin-Peierls Systems}
\def\runauthor{Hideo {\sc Yoshioka} and Yoshikazu {\sc Suzumura}}

\def\d{{\rm d}}

\def\dt{{\delta \theta}}
\def\on{{\omega_n}}
\def\dt2{(\delta \theta)^2}

\def\et{{\it et al.} }

\def\p{\sigma}
\def\al{\alpha}
\def\be{\beta}
\def\ga{\gamma}
\def\de{\delta}

\def\im{{\rm i}}

\def\lan{\left\langle}
\def\ran{\right\rangle}

\def\delx{\partial_x}

\def\deltau{\partial_\tau}

\def\nonum{\nonumber}

\def\dt{\frac{\partial}{\partial T}}

\def\e{{\rm e}}

\def\sgn{{\rm sgn}}

\def\e{{\rm e }}
\def\lsim{\lower -0.3ex \hbox{$<$} \kern -0.75em \lower 0.7ex \hbox{$\sim$}}
\def\gsim{\lower -0.3ex \hbox{$>$} \kern -0.75em \lower 0.7ex \hbox{$\sim$}}
\def\yen{Y \kern -1.077em =}
\def\jo #1#2#3#4{#1 {\bf #2} (#3) #4}   
\def\PR{Phys.\ Rev.}
\def\PRB{Phys.\ Rev.\ B}
\def\PRL{Phys.\ Rev.\ Lett.}

\def\JPCON{J.\ Phys.\ Condens. Matter}

\def\JPSJ{J.\ Phys.\ Soc.\ Jpn.}

\def\ANP{Ann.\ Phys.}

\def\PHB{Physica B}

\def\EPL{Europhys.\ Lett}

\title
{
Spin Defects in Spin-Peierls Systems
}

\author
{
Hideo {\sc Yoshioka}
\footnote{E-mail: h44770a@nucc.cc.nagoya-u.ac.jp}
and Yoshikazu {\sc Suzumura} 
\footnote{E-mail: e43428a@nucc.cc.nagoya-u.ac.jp}
\\
}

\inst
{
Department of Physics, Nagoya University, \\
Nagoya 464-01
} 

\recdate{\today}

\abst{
 We examine  spin-Peierls systems  in the presence of spin  defects 
which are introduced by replacing  
magnetic ions $\rm{Cu^{2+}}$ 
with non-magnetic ones $\rm{Zn^{2+}}$ 
in ${\rm CuGeO_3}$. 
 By using  the action for the bosonized Hamiltonian, 
  it is shown directly  that the  antiferromagnetic state 
    induced by the spin defects coexists with 
     the   spin-Peierls states.  
 Further the doping dependences of both  transition temperature 
of spin-Peierls state and  the spin gap have been calculated. 
  The transition temperature of the 
  present estimation 
  shows good agreement quantitatively with that observed in 
${\rm Cu}_{1-\de} {\rm Zn}_\de {\rm O}_3$ for the region of the 
doping rate,  $\de<0.02$. 
}

\kword{spin-Peierls state, impurity effect, ${\rm CuGeO_3}$}

\begin{document}
\sloppy
\maketitle
\section{Introduction}
The discovery of inorganic spin-Peierls (SP) system ${\rm CuGeO_3}$, 
which has chains of ${\rm Cu^{2+}}$ with spin $1/2$ 
along $c$-axis\cite{Hase-1}, 
 has attracted much attention to the role of disorder in SP systems. 
 In the material, the antiferromagnetic (AF) exchange energy 
 between nearest neighbor spins  is estimated as  $J = 183 K$ 
from  the magnetic field which corresponds  to saturation of 
magnetization\cite{Nojiri}. 
 The compound in the absence of disorder undergoes 
  SP transition at $T_{SP} = 14 K$\cite{Hase-1}, below which 
  the lattices spontaneously dimerize   
and the magnetic susceptibility rapidly drops to zero along all 
 the axes 
owing to appearance of the energy gap between the singlet ground state and 
the triplet excited one. 
The spin gap is estimated as $\Delta_S = 2.1 meV$ 
from the inelastic neutron scattering\cite{Nishi}.   

The effects of the disorder on SP systems 
have been mainly investigated on the materials where  
 Cu ions are  replaced by 
Zn\cite{Hase-2,Hase-3,Oseroff,Lussier,Hase-4,Sasago,Martin} or 
Ge ions are replaced by Si\cite{Renard,Poirier,Regnault}. 
It has been observed  that the doping decreases $T_{SP}$. 
Further, 
the fact that a new AF ordered phase appears 
below $T_{SP}$ 
  has been now well established 
 from measurements of the magnetic susceptibility 
 and the neutron scattering. 
This fact shows that the AF states coexists with 
 the SP state.  

 In terms of   the phase representation of SP Hamiltonian,
 Fukuyama \et 
  examined  the states between the two impurities 
 by assuming  the boundary condition that  the lattice dimerization is suppressed
at the location of the impurities 
\cite{Fukuyama}. 
 Treating  classical equations 
 with renormalized quantum fluctuation, 
 they have shown  that 
the AF state coexists with  the SP state 
in the presence of disorder. 

In the present paper, we investigate the effect of the spin defects 
  on SP states without the  assumption for the boundary condition 
 by deriving  the action for the SP system 
 and  show the coexistence of the two kinds 
of states.  
 It turns out that our results  are consistent with 
  those given by Fukuyama \et.    
 Further, we examine the variation  of both $T_{SP}$ and 
  $\Delta_S$ by the increase of the doping rate. 
In \S 2, we derive the action of the SP systems with the spin defects 
from Peierls-Hubbard model. 
The transition temperature is calculated in \S 3. 
In \S 4, we prove the coexistence of AF state and SP one and 
investigate the doping dependence of $\Delta_S$. 
Section 5 is devoted to summary and discussion.       

\section{Model}
 Since the replacement of ${\rm Cu}$ with ${\rm Zn}$ 
  gives rise to the variation of  
 both the charge and spin degrees of freedom,  
   we examine the effects of the spin defects  
 by starting the  Peierls-Hubbard model in one-dimension at half filling    
instead of the conventional SP model.  
 The Hamiltonian is given  as
\begin{eqnarray}
{\cal H} &=& 
-t \sum_{l,\p} 
\left\{ 1 + \lambda (u_l - u_{l+1}) \right\} 
\left( c^\dagger_{l,\p} c_{l+1,\p} + h.c. \right)
\nonum \\
&+& U \sum_l c^\dagger_{l,\uparrow} c_{l,\uparrow} 
c^\dagger_{l,\downarrow} c_{l,\downarrow}
+ \frac{K}{2} \sum_l (u_l - u_{l+1})^2, 
\label{eqn:PHM} 
\end{eqnarray}
where $t$ and $U$ are the hopping and 
the on-site Coulomb energy, respectively.   
 The quantity $\lambda$ is the coupling constant of the 
electron-lattice interaction and  
$c^\dagger_{l,\p}$ is a creation operator of electrons with the site $l$ and the spin $\p$.  The quantity $K$ is 
 the elastic constant with  
  $u_l$ being  the lattice distortion at  the site $l$. 

The model in eq.(\ref{eqn:PHM})
can be expressed by the phase variables as follows\cite{Fukuyama-Takayama},
\begin{eqnarray}
{\cal H} &=& 
\frac{v_\rho}{4 \pi} \int \d x 
\left\{
\frac{1}{K_\rho} (\delx \theta_{+})^2 + K_\rho (\delx \theta_{-})^2
\right\} 
\nonum \\
&+& \frac{v_\rho g_\rho}{\pi \alpha^2} \int \d x \cos 2 \theta_{+}
\nonum \\
&+&
\frac{v_\p}{4 \pi} \int \d x 
\left\{
\frac{1}{K_\p} (\delx \phi_{+})^2 + K_\p (\delx \phi_{-})^2
\right\}
\nonum \\
&+& \frac{v_\p g_\p}{\pi \alpha^2} \int \d x \cos 2 \phi_{+}
\nonum \\
&+& B \int \d x u(x) \sin \theta_{+} \cos \phi_{+} 
+ \frac{2K}{a} \int \d x u^2(x),   
\label{eqn:phaseH}
\end{eqnarray}
where $a$ is the lattice constant, 
 $B \propto  \lambda$  
and $u(x) = (-1)^l u_l$ with $x=la$. 
 The quantity  $\al$ denotes the cut-off 
 for the large wavenumber.    
 The velocity of excitation of the charge (spin) is given by 
$v_\rho$ ($v_\p$). 
 The quantities $g_\rho$ and $g_\p$ are the coupling constants 
    of the interaction for the umklapp scattering and the backward 
   scattering  respectively.  
 The coefficients $K_\rho$ and $K_\p$ which also include 
  parameters of interaction   
   characterize one-dimensional system.  
The phase variables, $\theta_\pm$ and $\phi_\pm$, 
express fluctuations of 
the charge and the spin degrees of freedom, respectively, 
 and satisfy the commutation relations, 
$[\theta_+(x), \theta_-(x)] = [\phi_+(x), \phi_-(x)] = \im \pi \sgn(x-x')$ 
and the others are zero. 
Note that we assumed $t > U$ in deriving eq.(\ref{eqn:phaseH}).  
The density of the $z$ component of the spin is expressed by 
$\delx \phi_+ / (2 \pi) - (-1)^{l} \sin \theta_{+} \sin \phi_{+} / \pi \al$, 
where the first and the second term express the slowly varying part, $S_z^0$,  
and the staggered one, $S_z^{\pi}$, respectively.   
  
In the above phase Hamiltonian, the charge excitation has a gap and 
the phase variable $\theta_{+}$
is fixed to the value of $\theta_{+} = (n + 1/2)\pi$ ( $n$ : integer ). 
 Thus, 
 eq.(\ref{eqn:phaseH}) 
 in the absence of disorder 
 becomes the same as that of phase representation 
of the conventional SP Hamiltonian\cite{Nakano}. 
The fact indicates that the regime of 
 the  weak interaction  is 
 analytically connected to that of the strong interaction. 
 Therefore the parameters in eq.(\ref{eqn:phaseH}) can be read 
  as those of phase Hamiltonian of SP model. 
  
The defects of spins are introduced as follows. 
 At the location 
  where ${\rm Cu}^{2+}$ are replaced by ${\rm Zn}^{2+}$, 
 there appears the charge defect which  
 is  expressed by as a kink of    $\theta_+$. 
 Since the fluctuation of the charge density 
is given by $\delx \theta_{+}/\pi$,  
  the quantity $\theta_+$ jumps by $\pi$\cite{Fukuyama-2} at the defect  
where  $\sin \theta_+$ varies from $\pm 1$ to $\mp 1$ resulting in  
 the change of the sign of the coupling between spin and lattice. 
Thus the action of SP systems 
in the presence of the spin defects is obtained as 
\begin{eqnarray}
S &=& \frac{v_\p}{4\pi K_\p} \int \d \tau \d x 
\left\{ 
(\delx \phi_+)^2 + \frac{1}{v_\p^2} (\deltau \phi_+)^2
\right\}
\nonum \\
&+& \frac{v_\p g_\p}{\pi \alpha^2} \int \d \tau \d x \cos 2 \phi_{+}
\nonum \\
&+& B \int \d \tau  \d x m(x) u(x) \cos \phi_+
+ \be \frac{2K}{a} \int \d x u^2(x), 
\label{eqn:action}
\end{eqnarray}
where 
 $\be$ is the inverse of temperature, $T$.    
 The quantity $m(x)$, which satisfies $|m(x)| = 1$, 
  changes the sign at the location of the spin defects.  
  The staggered component of $S_z^\pi$ is given by 
$ - (-1)^{l} m(x) \sin \phi_{+}(x) / \pi \al$. 
In the following, we study the case 
where the spin systems are expressed by Heisenberg chains, 
where $v_\p = \pi J a / 2$\cite{velocity}, 
$K_\p \to 1$ due to the rotational symmetry in spin space\cite{Luther-Peschel} 
and $g_\p \to 0$\cite{weak}.  

\section{Temperature of Spin-Peierls Transition}

\begin{figure}
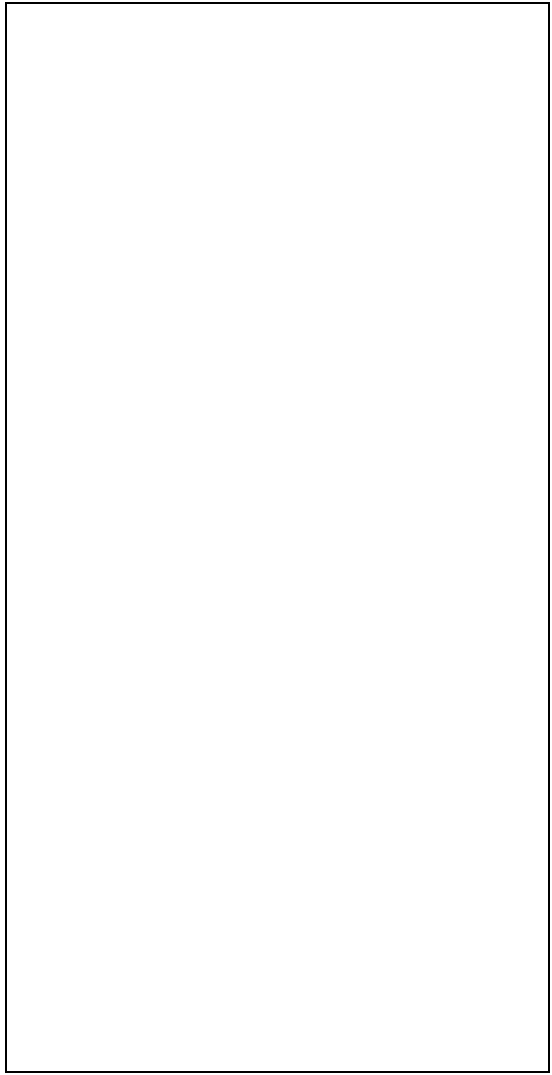

\begin{center}
\figureheight{14cm}
\end{center}
\caption{
(a) The critical temperature of spin-Peierls transition normalized by 
that without doping as a function of $(J/T_{SP}^0) \de$. 
The dotted (dashed) curve shows an asymptotic form for $J \de / T_{SP}^0 \ll 1$
($\gg 1$) given by eq.(\ref{eqn:TSPzenkin}).
(b) Comparison of the present result 
 with the experimental result (open and filled circles) 
  in ref.\citen{Martin}
 where  $J = 183K$ and $T_{SP}^0 = 14K$. 
Here the open and the filled circles expresses $T_{SP}$ 
determined by the measurements of 
the neutron scattering and the magnetic susceptibility, respectively.   
}
\label{fig:TSP}
\end{figure}

 We calculate $T_{SP}$ from the softening of phonon\cite{Cross} 
 with $q = \pi$. 
 The complete softening takes place at $T = T_{SP}$   
 resulting in the  vanishing of the effective elastic constant, 
 $K_{eff}$, which comes from the 
 renormalization of $K$ in eq.(\ref{eqn:action}) through 
 the electron-phonon coupling.   
By integrated out the spin degree of freedom up to the second order,  
$K_{eff}$ is given as follows, 
\begin{equation}
K_{eff} = K - \frac{a}{\omega_c} B^2 \int_0^\infty \d x 
h(x) {\rm e}^{- x/\xi} K({\rm e}^{-2x/\xi}), 
\label{eqn:elastic}  
\end{equation}
where $\omega_c = v_\p \al^{-1}$, $\xi = v_\p / (\pi T)$ and 
$K(k)$ is the complete elliptic integral of the first kind. 
The function $h(x)$ is defined  by $(1/L)\int \d X \lan m(X+x/2) m(X-x/2)\ran_{imp}$ 
with $\lan \cdots \ran_{imp}$ expressing the impurity average.  
 By noting that  $h(x)$ is given by 
  $1 - 2 \de |x|/a$  for $ 2 \de |x|/a \ll 1$ and  
 $h(x \to \pm \infty) = 0$, 
we make use of the approximation 
 given by  $e^{-2\de |x|/a}$ where $\de$ denotes the doping rate.   
 Then $T_{SP}/T_{SP}^0$ with $T_{SP}^0$ being the transition temperature 
in the absence of the impurities is determined self-consistently 
by the following equation, 
\begin{equation}
\frac{T_{SP}}{T_{SP}^0} = 
\frac{1}{C_1} \int_0^1 \d u u^{(J \de /T_{SP}^0)/(T_{SP}/T_{SP}^0)} K(u^2), 
\label{eqn:TSP}
\end{equation}
where $C_1 \equiv \int_0^1 \d t K(t^2) \simeq 1.72$. 
The asymptotic behavior is given as follows, 
\begin{eqnarray}
& &\frac{T_{SP}}{T_{SP}^0} 
\nonum \\
&\simeq& \left\{
\begin{array}{ll}
\displaystyle{
1 - \frac{C_2}{C_1} \frac{J \de}{T_{SP}^0} } 
&
\displaystyle{ 
{\rm for} \hspace{1em} \frac{J \de}{T_{SP}^0} \ll 1
}
\\
\displaystyle
{ 
4 \frac{J \de}{T_{SP}^0}
\exp\left\{
- \left(2 C_1 \frac{J \de}{T_{SP}^0} - \ga \right)
\right\} 
}
& 
\displaystyle{ 
{\rm for} \hspace{1em} \frac{J \de}{T_{SP}^0} \gg 1
}
\end{array}, 
\right. 
\nonum \\
& &
\label{eqn:TSPzenkin}
\end{eqnarray}
where $C_2 \equiv - \int_0^1 \d t \ln t K(t^2) \simeq 1.59$ and 
$\ga \simeq 0.577$ is Euler's constant.   
We show $T_{SP}/T_{SP}^0$ as a function of $J \de / T_{SP}^0$ 
in Fig.\ref{fig:TSP}(a). 
According to the measurements of magnetic susceptibility
 and neutron scattering, 
$T_{SP}$ of ${\rm Cu}_{1-\de} {\rm Zn}_\de {\rm Ge} {\rm O}_3$ 
decreases linearly for $\delta < 0.02$ and saturates 
from $\delta \simeq 0.02$ to at least 
$\delta \simeq 0.05$\cite{Sasago,Martin}. 
In Fig.\ref{fig:TSP}(b), 
 by substituting  the values of $J=183K$ and $T_{SP}^0 = 14K$  
 into eqs.(\ref{eqn:TSP}) and (\ref{eqn:TSPzenkin}),  
 the present result 
 (solid curve and dashed line) is compared 
 with the experimental one given by the measurements of 
the neutron scattering (open circles) 
 and the magnetic susceptibility (filled circles) which is  
 shown in Fig.\ref{fig:TSP} of ref.\citen{Martin}.  
 Figure \ref{fig:TSP}(b) shows  good agreement between 
  the present result and the  experimental one for $\delta \lsim 0.02$. 
 However, our result 
 cannot explain the saturation of the transition temperature  
  in the region of the doping rate of  $0.02 < \de < 0.05$. 

\section{The Case of $T \ll T_{SP}$}

In this section, we examine the coexistence of the AF state and 
 the SP state in the presence of the spin defects 
and calculate the doping dependence of $\Delta_S$
based on the action given by  eq.(\ref{eqn:action}).  

  We note that the change of the sign of the spin-lattice interaction 
 at the location of the spin defects 
 can be reexpressed as 
 $ m(x) \cos\phi_{+}(x) \to  
\cos \left\{ \phi_{+}(x) + n(x) \pi \right\}$
and
$ m(x) \sin \phi_{+}(x) \to  
\sin \left\{ \phi_{+}(x) + n(x) \pi \right\}$ 
where the integer $n(x)$  jumps alternatively by $\pm 1$ 
at  the spin defects. 
By rewriting $\phi_{+} + n(x) \pi$ as  $\phi$, 
the action, $S$, can be given as follows, 
\begin{eqnarray}
S &=& \frac{v_\p}{4\pi} \int \d \tau \d x 
\left\{ 
(\delx \phi)^2 + \frac{1}{v_\p^2} (\deltau \phi)^2
\right\}
\nonum \\
&-& \frac{v_\p}{2}  \int \d \tau \sum_i p_i \delx \phi(x_i) 
+ \be \frac{\pi v_\p}{4} \sum_{i,j} p_i p_j \de(x_i - x_j) 
\nonum \\
&+& B \int \d \tau  \d x u(x) \cos \phi
+ \be \frac{2K}{a} \int \d x u^2(x), 
\label{eqn:action2}
\end{eqnarray}
where 
$x_i$ is the position of the spin defects. 
 With the use of 
 the number $p_i = \pm 1$,  
$\delx n(x) = \sum_i p_i \de(x-x_i)$ 
 and  $p_i p_{i+1} = -1$. 
In eq.(\ref{eqn:action2}), 
we divide the lattice distortion as 
$u(x) = - u + \bar{u}(x)$
where   
 $\bar{u}(x)$ denotes the spatial variation around 
  the uniform value, $u$.  
Correspondingly, the action $S$ is divided as 
$S = S_1[u] + S_2[u,\bar{u}(x)]$. 
We apply the variational method to $S_1[u]$,  
whose trial action $S_0[u]$ is given by
\begin{eqnarray}
S_0[u] &=& \frac{v_\p}{4\pi} \int \d \tau \d x 
\left\{ 
(\delx \phi)^2 + q_0^2 \phi^2 +  \frac{1}{v_\p^2} (\deltau \phi)^2
\right\}
\nonum \\
&-& \frac{v_\p}{2}  \int \d \tau \sum_i p_i \delx \phi(x_i)
\nonum \\ 
&+& \be \frac{\pi v_\p}{4} \sum_{i,j} p_i p_j \de(x_i - x_j) 
+ \be L \frac{2K}{a} u^2. 
\label{eqn:t-action}
\end{eqnarray}
Here $L$ is the system size, and $q_0$ is related to 
the spin gap in the presence of the disorder $\Delta_S$ 
as $\Delta_S = v_\p q_0$.  
The quantities, $q_0$ and $u$ are determined by minimizing the following 
Free energy\cite{Feynman},     
\begin{equation}
F_{tr} = F_0 
- B u \int \d x \lan \cos \phi \ran_{S_0}
- \frac{v_\p q_0^2}{4 \pi} \int \d x \lan \phi^2  \ran_{S_0}, 
\label{eqn:f-energy}
\end{equation}
where $\lan \cdots \ran_{S_0}$ expresses the average 
with respect to $S_0 [u]$. 
On the other hand, ${\bar u}(x)$ 
is determined by minimizing the quantity,  
$ - \ln \lan \exp \left\{- S_2 [u, {\bar u(x)}] \right\} \ran_{S_0}$ 
with respect to ${\bar u}(x)$.    

In order to calculate the various quantities averaged by $S_0$, 
we first calculate the generating function, 
$W[J(x,\tau)] \equiv 
\lan 
\exp \left\{- \im \int \d \tau \d x J (x,\tau) \phi(x,\tau) \right\}
\ran_{S_0}
$ with $J(x,\tau)$ being an arbitrary real function.  
By applying path integral method, the generating function 
is given as follows, 
\begin{eqnarray}
W[J(x,\tau)] &=& \exp 
\Big\{
- \frac{1}{2} \int \d \tau \d x \d \tau' \d x'
\nonum \\ 
& & H(x-x', \tau-\tau';T) J(x,\tau) J(x',\tau')  
\Big\}
\nonum \\
&\times& 
\exp 
\left\{
- \im \int \d \tau \d x F(x) J(x,\tau) 
\right\},
\label{eqn:g-func}  
\end{eqnarray} 
where 
\begin{eqnarray}
& & H(x-x', \tau-\tau';T) 
\nonum \\
&=& \frac{1}{\be L} \sum_{q,\on}
\frac{
\exp \left\{ - \im q (x-x') + \im \on (\tau-\tau') \right\}
}
{
\displaystyle{\frac{v_\p}{2 \pi}} ( q^2 + q_0^2 + \on^2 )
}, 
\label{eqn:H-func}
\end{eqnarray}
and
\begin{eqnarray}
F(x) 
&=& \frac{\pi}{2} \sum_i p_i \frac{\pi}{L} \sum_q
\frac{\im q \exp \left\{ \im q (x_i - x) \right\} }
{q^2 + q_0^2}
\nonum \\
&=& \frac{\pi}{2} \sum_i p_i 
\e^{
- q_0 |x - x_i|
}
\sgn(x-x_i),
\label{eqn:F-func}
\end{eqnarray}
with $\on = 2 n \pi T$ ( $n$ : integer ).
Equations (\ref{eqn:g-func})-(\ref{eqn:F-func}) 
lead the following results, 
\begin{eqnarray}
S_z^\pi(x) &=& 
- \frac{(-1)^l}{\pi \al}  
\lan \sin \phi \ran_{S_0} \nonum \\
&=& 
- \frac{(-1)^l}{\pi \al}  
\exp\left\{ - \frac{1}{2} H (0,0;T) \right\} \sin F(x),
\label{eqn:SZ} 
\\
u(x) &=& 
-  \frac{aB}{4K} 
\lan \cos \phi \ran_{S_0} \nonum \\ 
&=& 
- \frac{aB}{4K} 
\exp\left\{ - \frac{1}{2} H (0,0;T) \right\} \cos F(x). 
\label{eqn:cos} 
\end{eqnarray}
with $x = la$. 
Here the first line in eq.(\ref{eqn:cos}) is derived 
by minimizing   
$ - \ln \lan \exp \left\{- S_2 [u, {\bar u(x)}] \right\} \ran_{S_0}$ 
 with respect to ${\bar u}(x)$. 
 The function $|F(x)|$ takes almost a value of $\pi/2$ 
 near the location of the spin defects and 
  decreases rapidly  to zero far from the location. 
Thus, without the assumption of the boundary condition,  
we obtain that 
the lattice dimerization is suppressed and AF moments appear 
near the location of the spin defects. 
The characteristic length 
of the spatial dependence is given by $1/q_0$. 
\begin{figure}
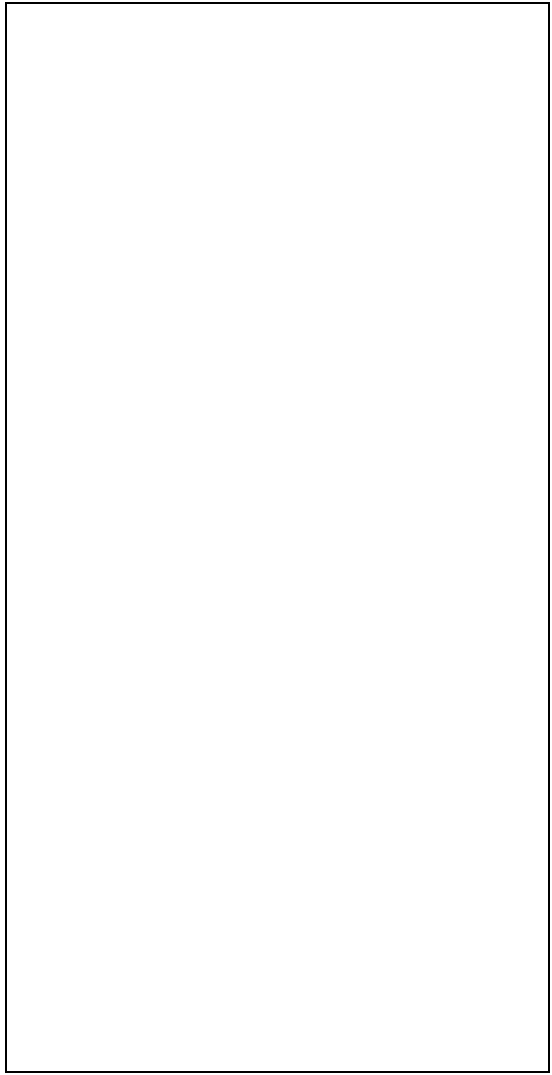

\begin{center}
\figureheight{14cm}
\end{center} 
\caption{
Spatial dependence of the classical part of the phase, $\theta_{cl}$, 
in the case of two impurities, 
where the distance between the two impurities are chosen as $159a$(a)
and $79a$(b) with $a$ being the lattice distance. 
The solid ( dotted ) curves  express the classical part of the phase obtained from  the present study 
  ( the classical equation in ref.\citen{Fukuyama}).     
}
\label{fig:compare}
\end{figure}
Equations (\ref{eqn:SZ}) and (\ref{eqn:cos}) show the facts that 
the phase $\phi(x)$ can be divided as $\phi(x) = \phi_{cl}(x) + \phi_q(x)$ 
where $\phi_{cl}(x)$ is the classical part and  
$\phi_q(x)$ expresses the quantum fluctuation around $\phi_{cl}(x)$ where   
$\phi_{cl}(x) = F(x)$  and $\lan \phi_q^2(x) \ran = H(0,0;T)$. 
In Fig.\ref{fig:compare},  
we compare the classical part of the phase obtained in the present study 
 and that derived by solving the classical equation\cite{Fukuyama} 
in the case of two defects.
Here the distance between the impurities, $l_{imp}$ 
is $159a$ (a) and $79a$ (b) 
and $1/(q_0 a)$ is chosen as 11.8 which corresponds to undoped 
${\rm CuGeO}_3$. 
The solid curve expresses 
$\theta_{cl}(x) = F(x) + \pi/2$ and the dotted curve is obtained 
from the classical equations 
with the boundary conditions of $\theta_{cl}(0) = \theta_{cl}(l_{imp}) =
F(0) + \pi/2$. 
Note that we have chosen the region of $F(0) = F(l_{imp}) \simeq - \pi/2$. 
Figure \ref{fig:compare} shows the following facts. 
Our treatment that the cosine term is replaced by the term proportional 
to $\phi^2$ underestimates the correlation between the impurities. 
Our result is valid for the case of the small doping, 
$q_0 a / \de \gg 1$. 

The free energy of eq.(\ref{eqn:f-energy}) per unit length, 
$f_{tr} \equiv F_{tr}/L$ 
is easily calculated as
\begin{eqnarray}
f_{tr} &=& \frac{1}{L} \sum_q \frac{E_q}{2}
+ \frac{1}{\be L} \sum_q \ln ( 1 - e^{- \be E_q} ) 
\nonum \\
&+& \frac{1}{L} \frac{\pi v_\p q_0}{16}
\sum_{ij} p_i p_j 
\left( 1 + q_0 |x_i-x_j|  \right)
\e^{-q_0|x_i - x_j|}
\nonum \\ 
&+& \frac{2K}{a} u^2
- B u  \e^{- \frac{1}{2} H(0,0;T)}
  \frac{1}{L} \int \d x \cos F(x) 
\nonum \\  
&-& \frac{v_\p q_0^2}{4\pi} H(0,0;T), 
\label{eqn:ftr}
\end{eqnarray}
where $E_q = v_\p \sqrt{q^2 + q_0^2}$ is the excitation spectrum of spin 
fluctuation. 
In the following, we consider the case of $T=0$ and 
$q_0 a /\de \gg 1$. 
Then the energy per unit length, $\epsilon_{tr} \equiv f_{tr}(T=0)$, 
can be calculated as follows,
\begin{eqnarray}
\epsilon_{tr} 
&=& \frac{v_\p}{2 \pi \al^2} \left( \frac{\al q_0}{2} \right)^2 
+ \frac{v_\p}{\pi \al^2} \frac{\al \pi^2}{8a} \de 
\left( \frac{\al q_0}{2} \right) 
+ \frac{2K}{a} u^2 
\nonum \\
&-& B u \left( \frac{\al q_0}{2} \right)^{1/2}
\left\{
1 - \frac{\al}{a(\al q_0/2)} Cin(\frac{\pi}{2}) \de
\right\}, 
\label{eqn:energy}
\end{eqnarray}
where $Cin(\pi/2) = \int_0^{\pi/2} \d t ( 1 - \cos t )/t \simeq 0.5568$
\cite{suuchi}. 
In deriving eq.(\ref{eqn:energy}), we used $q_0 \al \ll 1$. 
By minimizing eq.(\ref{eqn:energy}) with respect to $u$ and $\al q_0/2$,  
 and defining  $\Delta_S^0$ and $u_0$ as 
the spin gap and the lattice dimerization in the absence of the disorder, 
 quantities $\Delta_S/\Delta_S^0$ and $u/u_0$ 
in the case of $J \de/\Delta_S \ll 1$ 
 are given as follows, 
\begin{eqnarray}
\frac{\Delta_S}{\Delta_S^0} &=& 
1 - \frac{\pi^2}{8} \left( \frac{J\pi}{\Delta_S^0} \right) \de, 
\label{eqn:gap}
\\
\frac{u}{u_0} &=& 1 - 
\left\{
\frac{\pi^2}{16} + Cin(\frac{\pi}{2})
\right\}
\left( \frac{J\pi}{\Delta_S^0} \right) \de . 
\label{eqn:dimerization} 
\end{eqnarray}
Note that eq.(\ref{eqn:dimerization}) is consistent with 
eq.(\ref{eqn:cos}) in the sense that the spatial average 
of eq.(\ref{eqn:cos}) leads to eq.(\ref{eqn:dimerization}). 
Thus the spin gap and the lattice dimerization are suppressed linearly
as a function of doping as is seen in the transition temperature. 
The ratio, $\Delta_S/T_{SP}^0$, is given as 
\begin{equation}
\frac{\Delta_S}{T_{SP}} =  
\frac{\Delta_S^0}{T_{SP}^0} 
\left\{
1 - 
\left(
\frac{\pi^2}{8} \frac{\pi J}{\Delta_S^0} - \frac{C_2}{C_1}\frac{J}{T_{SP}^0}
\right) \de 
\right\} .  
\label{ratio}  
\end{equation}
 For the case of 
 ${\rm Cu}_{1-\de}{\rm Zn}_\de {\rm Ge} {\rm O_3}$,
 the quantity $\Delta_S/T_{SP}$ 
 decreases with increase of the doping rate, 
 i.e.,  $\Delta_S/T_{SP} \simeq \Delta^0_S/T^0_{SP}(1-17\de)$. 
The decrease of the spin gap originates from the fact that
  the coupling between the spins decreases 
 due to the suppression of the lattice dimerization around
   the spin defects. 
 Such a   result indicates that  
  the model having 
the  constant lattice dimerization with the value 
in the absence of the 
impurities\cite{Martins}  
 is hard to explain   
 the suppression of the spin gap. 

\section{Summary and Discussion}

  We  have derived  the action for the SP system 
 in the presence of the spin defects, and  
 have shown 
  that the lattice dimerization is suppressed 
and the AF moments appear around the location of the defects.  
The doping dependences of both the transition temperature 
  and the spin gap 
  were   calculated based on the action. 

 The action was obtained from the  Peierls-Hubbard model 
 which includes not only spin but also charge degrees of freedom,   
because the substitution  
 of non-magnetic ion for magnetic ion  influences 
both  degrees of freedom. 
The resulting action given in eq.(\ref{eqn:action}) 
is equal to that of SP model with spin-lattice coupling 
 which changes the sign at the location of the defects 
  due to the kink of the charge. 

The transition temperature of the present calculation 
  shows good agreement 
with the experimental result of 
${\rm Cu}_{1-\de}{\rm Zn}_\de {\rm Ge} {\rm O_3}$  
for $\delta < 0.02$. 
 However there is a discrepancy in the sense  that  
 the saturation of $T_{SP}$ appears  
 in the experiments  for $\delta > 0.02$ 
 while $T_{SP}$ of the present calculation 
 decreases exponentially by the increase 
  of $\delta$. 
  Further theoretical studies are needed in order to understand 
   the saturation of the transition temperature 
 of the SP state under doping. 

 By applying the variational method to the action, 
 the coexistence of the AF state and the SP one 
 at  $T=0$   was demonstrated  without the assumption of the boundary condition at the spin defect. 
The present results  are consistent with the assumption 
and the result by Fukuyama \et.  
In addition, it was shown that 
the spin gap and the spatially averaged lattice dimerization
decrease by the doping. 
The decrease of the spin gap is 
due to the suppression of the lattice dimerization near the spin defects. 
  
Finally we comment on the low energy excitation in the spin 
gap\cite{Hase-4,Sasago,Martin,Regnault,Martins,Saito}.
It is very complicated  to discuss the excitation by the variational method. 
 However the existence of the excitation 
  in the present action 
 of eq.(\ref{eqn:action}) could be understood qualitatively   
  by the following discussion. 
 According to the renormalization group analysis, 
the coefficient $B$ in eq.(\ref{eqn:action}) tends to strong coupling 
for $K_\p < 4$, which shows that 
the system described by eq.(\ref{eqn:action}) with $K_\p < 4$ 
belongs to the same universality class.   
In the special case of $K_\p = 2$\cite{Luther-Emery}, 
the Hamiltonian of the spin 
sector corresponding 
to eq.(\ref{eqn:action}) can be  mapped into 
 that of  non-interacting Dirac Fermions 
systems with the mass, $m_0(x)$,    
\begin{eqnarray}
\cal{H} &=& v_\p \int \d x 
\left\{
\psi_1^\dagger (-\im \delx)\psi_1 - \psi_2^\dagger (-\im \delx)\psi_2 
\right\}  
\nonum \\
&+& \int \d x 
\left\{ m_0(x)
\left( \psi_1^\dagger \psi_2 + \psi_2^\dagger \psi_1  \right)
\right\}, 
\label{eqn:HK2}  
\end{eqnarray}
where $m_0(x) = \pi \al B m(x) u(x)$. 
 The term proportional to
$\cos 2 \phi_+$ is neglected because the term is less divergent 
compared to that proportional to $\cos \phi_+$. 
In the case where $m_0(x)$ takes periodically two values 
$\phi_0$ and $-\phi_0$ with the same interval $l_{int}$,
 the gap near the zero energy vanishes, 
because $m_0(q=0) = \int_0^{2l_{int}} \d x m(x) = 0$. 
  We also  note another  case 
 that  $m_0(x)$ takes two values $\phi_0$ and $\phi_1$ alternatively   
with the random distribution of 
 the interval length $l$. 
  When  the distribution is given by 
  $f_0(l) = \theta(l) n_0 \e^{-n_0l}$ and 
$f_1(l) = \theta(l) n_1 \e^{-n_1l}$, 
 i.e.,  $1/n_0$ and $1/n_1$ being the mean length 
 for $\phi_0$ and $\phi_1$ respectively, 
 the integrated density of states and the localization length can be 
 calculated exactly \cite{Comtet}. 
 Fabrizio and M\'elin\cite{Fabrizio} have noticed that 
the Hamiltonian of SP system of XY chain with doping of ${\rm Zn}$ 
is given by eq.(\ref{eqn:HK2})  
and showed the midgap states 
by applying the above method. 
 For the quantitative discussion, 
 one must take into account the fact 
 that the spatial variation of the mass, $m_0(x)$, 
 is not exactly the telegraph type but  changes the sign with 
 the characteristic length of the order of $1/q_0$
  due to  $u(x) \propto \cos F(x)$,  
 and that  the 
 magnitude of the mass depends on the doping rate.  
The further exploration is needed in order to understand
the low energy excitation in the presence of spin defects.

\section*{Acknowledgements}
This work is financially supported by 
a Grant-in-Aid for Scientific Research from the Ministry 
of Education, Science and Culture.


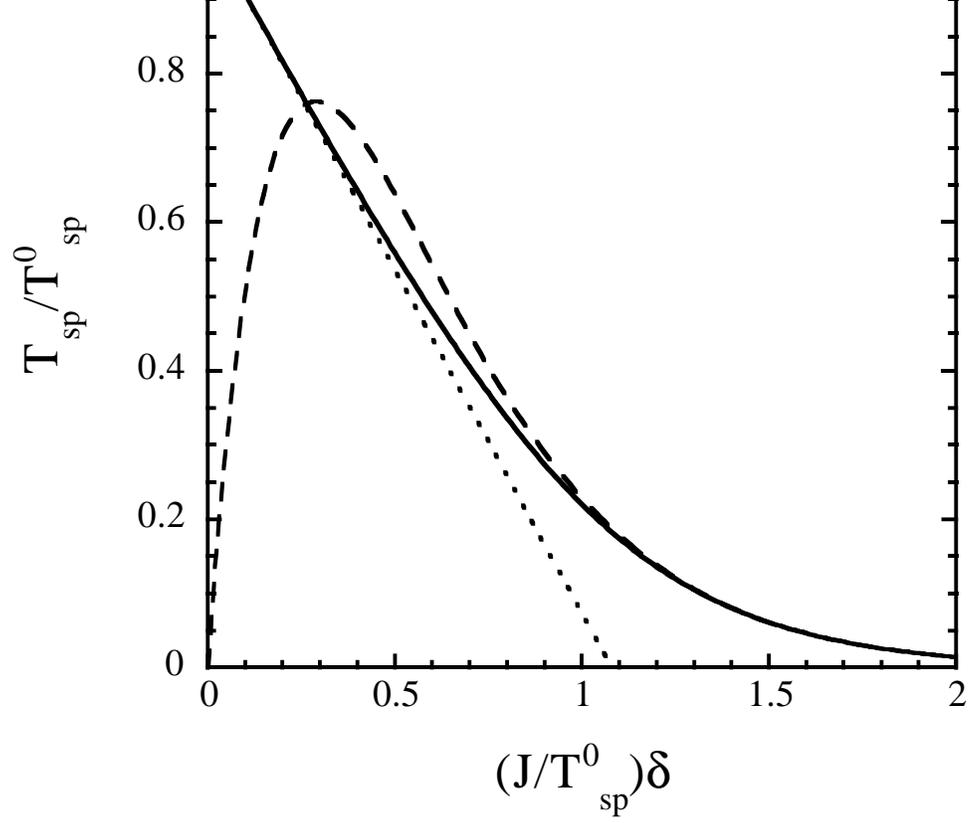
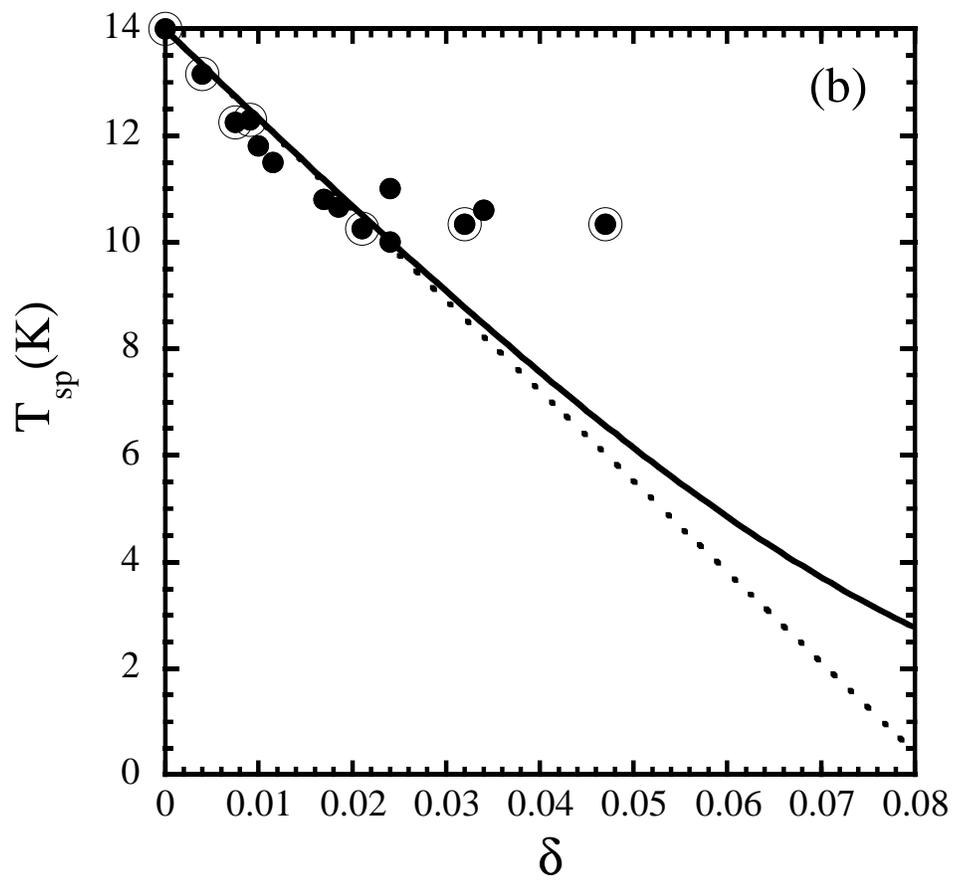
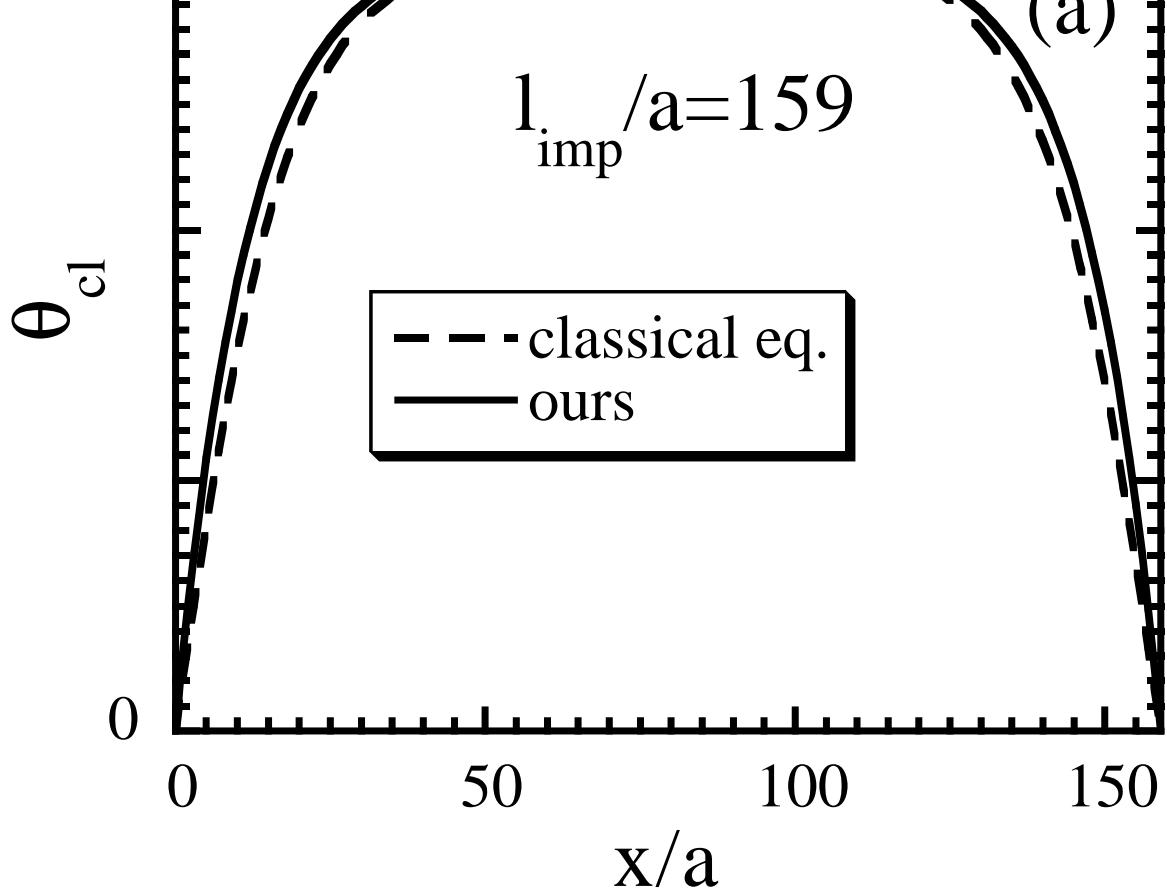
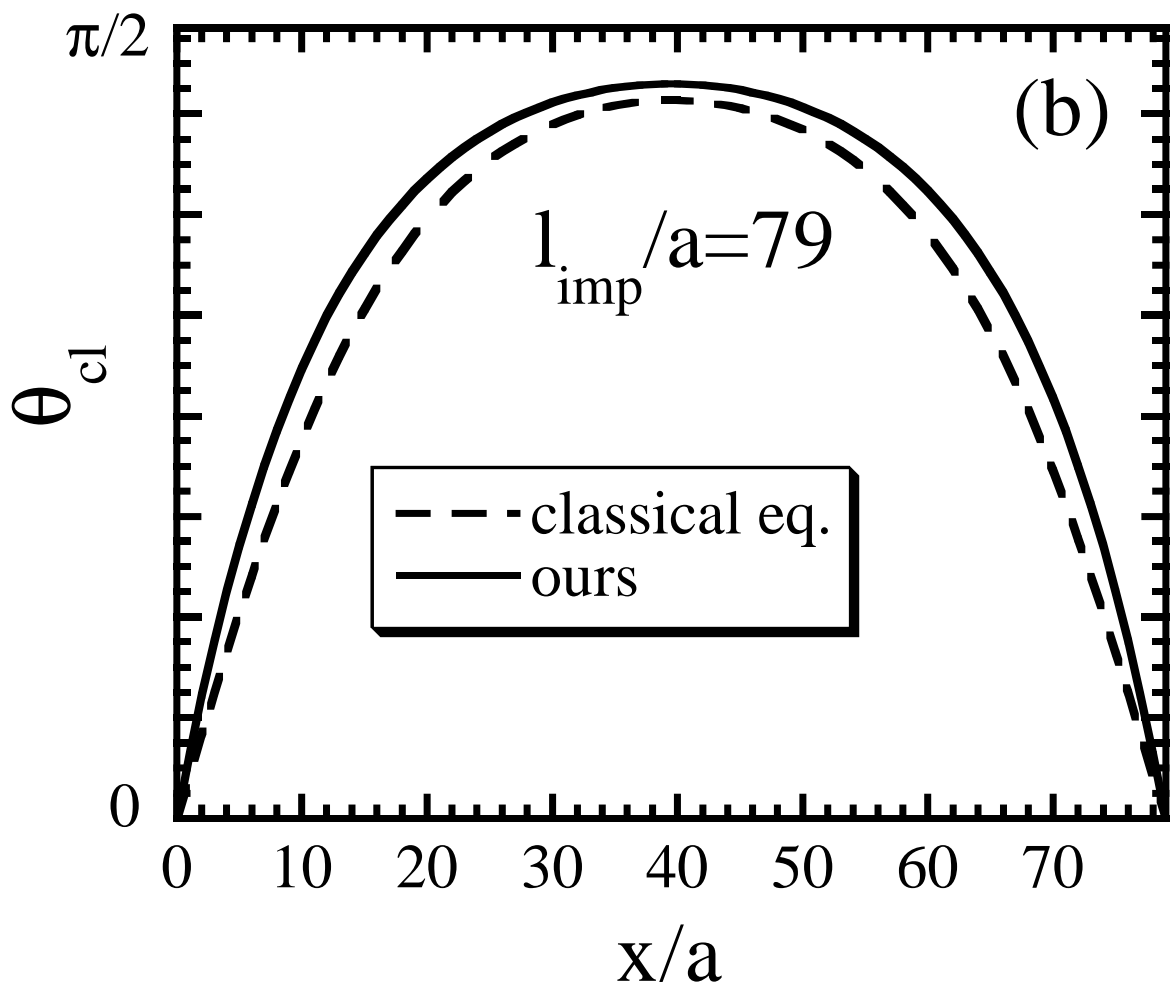
\end{document}